\documentclass[preprint]{elsarticle}
 
\usepackage{lineno,hyperref}
\usepackage{amsmath}
\usepackage{subcaption}
\usepackage{textgreek}
\usepackage{booktabs}
\usepackage{float}
\usepackage{mathtools}

\journal{JMBBM}

\begin{document}

\begin{frontmatter}

\title{A microstructural model of tendon failure}

\author[1]{James Gregory%
	 \corref{cor1}}
\ead{james.gregory@manchester.ac.uk}

\author[1,2]{Tom Shearer}

\author[1]{Andrew L. Hazel}

\address[1]{Department of Mathematics, University of Manchester, Manchester, UK}
\address[2]{Department of Materials, University of Manchester, Manchester, UK}

\cortext[cor1]{Corresponding author}

\begin{abstract}
Collagen fibrils are the most important structural component of tendons. Their crimped structure and parallel arrangement within the tendon lead to a distinctive non-linear stress-strain curve when a tendon is stretched. Microstructural models can be used to relate microscale collagen fibril mechanics to macroscale tendon mechanics, allowing us to identify the mechanisms behind each feature present in the stress-strain curve. Most models in the literature focus on the elastic behaviour of the tendon, and there are few which model beyond the elastic limit without introducing phenomenological parameters. We develop a model, built upon a collagen recruitment approach, that only contains microstructural parameters. We split the stress in the fibrils into elastic and plastic parts, and assume that the fibril yield stretch and rupture stretch are each described by a distribution function, rather than being single-valued. By changing the shapes of the distributions and their regions of overlap, we can produce macroscale tendon stress-strain curves that generate the full range of features observed experimentally, including those that could not be explained using existing models. These features include second linear regions occurring after the tendon has yielded, and step-like failure behaviour present after the stress has peaked. When we compare with an existing model, we find that our model reduces the average root mean squared error from 4.15MPa to 1.61MPa, and the resulting parameter values are closer to those found experimentally. Since our model contains only parameters that have a direct physical interpretation, it can be used to predict how processes such as ageing, disease, and injury affect the mechanical behaviour of tendons, provided we can quantify the effects of these processes on the microstructure. 
\end{abstract}


\end{frontmatter}


\section{Introduction}\label{section:introduction}
\begin{figure}[H]
	\centering
	\includegraphics{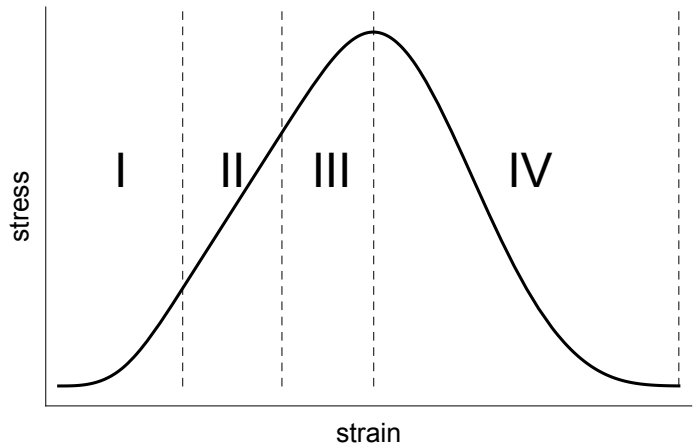}
	\caption{Idealised stress-strain behaviour of a tendon stretched to failure. Region I is the non-linear toe region, where collagen fibrils gradually become taut, increasing the overall tendon stiffness. Region II is the linear region, where all the fibrils are exhibiting a linearly elastic response. The end of region II is the macroscopic yield point, where yielding in the fibrils causes a reduction in gradient. Region III is the post-yield region, where fibrils begin to yield and fail. Region IV is the macroscopic failure region, where fibrils continue to fail and the whole tendon eventually ruptures.}
	\label{fig:introduction:tendon stress-strain}
\end{figure}

\begin{figure}[H]
	\centering
	\includegraphics{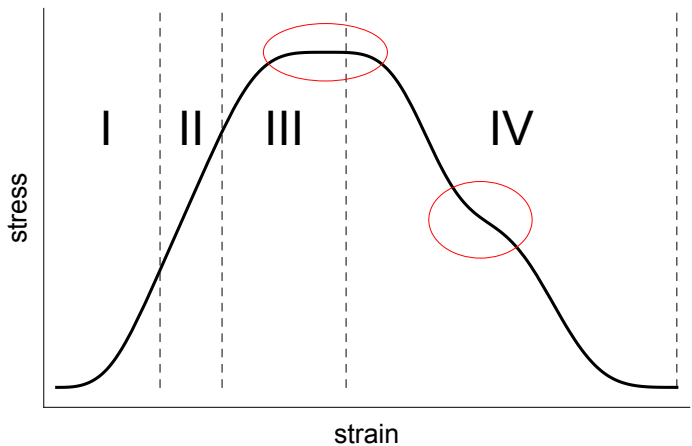}
	\caption{An idealised stress-strain curve demonstrating observed features (highlighted in red) which cannot be captured using existing models \cite{hurschler1997structurally,hamedzadeh2018constitutive}. Regions I and II are the same as in Figure \ref{fig:introduction:tendon stress-strain}, but the post-yield region (region III) shows a plateau instead of a well-defined peak. The macroscopic failure region (region IV) contains step-like failure behaviour, where the second derivative of the stress changes sign multiple times.}
	\label{fig:introduction: plateau and steps}
\end{figure}

Tendons are composed of a complex hierarchy of collagen-based components embedded within an extra-collagenous matrix. When modelling the mechanical response of tendons as they are stretched to failure, it is important to consider this complex microstructure and how it gives rise to the observed stress-strain behaviour illustrated in Figure \ref{fig:introduction:tendon stress-strain}. The macroscale tendon stress-strain curve can be split into four sections: I) the non-linear toe region, II) the linear region, III) the post-yield region, and IV) the macroscopic failure region. Existing microstructural models \cite{hurschler1997structurally,hamedzadeh2018constitutive} are able to capture this behaviour when it resembles the idealised case presented in Figure \ref{fig:introduction:tendon stress-strain}, but we will show that a significant proportion of observed stress-strain curves \cite{goh2018age} contain features that cannot be explained using these models. These features include second linear regions (in region III) and step-like failure behaviour (in region IV), as shown in Figure \ref{fig:introduction: plateau and steps}. We therefore propose a new microstructural model, based on the response of individual collagen fibrils, which is capable of capturing this behaviour.

Collagen fibrils often form the basis of microstructural models because they are the smallest component of tendons for which we have reliable stress-strain data. Fibrils are crimped within the fascicle, and only become load-bearing once the tendon has been stretched enough to remove their crimp. Due to varying crimp between fibrils, the tendon initially exhibits a non-linear stress-strain response, as fibrils gradually become taut (region I). The tendon stiffness continues to increase until all of the fibrils are taut and we see a macroscale linear region (region II). Hijazi et al. \cite{hijazi2019ultrastructural} used scanning electron microscopy to show that stretching a tendon past the end of its linear region results in permanent damage on the fibril scale, suggesting that as we pass into the post-yield region (region III), fibrils themselves begin to yield. When the tendon reaches the macroscopic failure region (region IV), many of the fibrils have yielded, and some may have ruptured completely. Eventually all of the fibrils will rupture and the tendon will fail.

	\begin{figure}[H]
	\centering
	\includegraphics{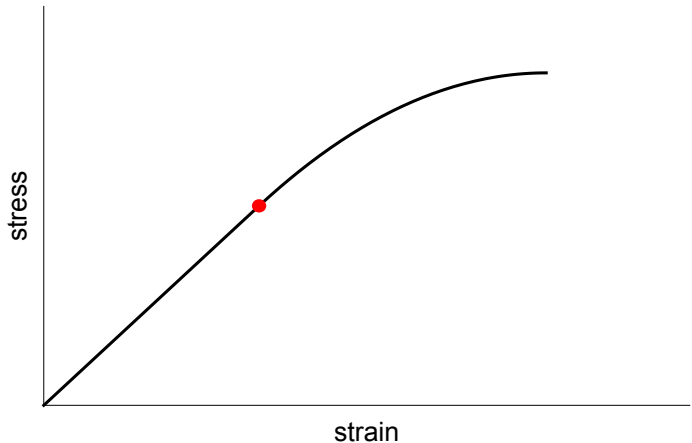}
	\caption{Idealised stress-strain behaviour of an individual collagen fibril stretched to failure. The fibril exhibits a linear response initially, before experiencing a decrease in gradient. Stretching beyond the linear region causes the fibril to become damaged \cite{shen2008stress}, leading us to refer to the transition between linear and non-linear behaviour as the fibril yield stretch/strain, represented above by the red marker.}
	\label{fig:introduction:fibril stress-strain}
	\end{figure}
 
To build a microstructural model capable of describing the full range of tendon stress-strain behaviour observed in regions I--IV, we must look in more detail at the mechanical response of individual collagen fibrils. Many research groups have performed failure tests on isolated fibrils \cite{shen2008stress, shen2010vitro, liu2016tension, yamamoto2017tensile, yamamoto2017relationships,svensson2013fracture}, and whilst the reported material parameters show a large amount of variability, several trends still emerge. The fibril stress-strain response is most often described as linear initially, before becoming non-linear with a decreasing slope  \cite{shen2008stress, shen2010vitro, liu2016tension, yamamoto2017tensile, yamamoto2017relationships}. Van der Rijt et al. \cite{van2006micromechanical} reported non-linear toe-regions for small strains ($<$4\%), but claim that the stress-strain curve appears to be almost perfectly linear when the fibril is stretched further. For this reason, we make the assumption that the non-linear toe-region is negligible and that the initial phase of the fibril stress is linear. Figure \ref{fig:introduction:fibril stress-strain} shows an idealised stress-strain curve for an isolated collagen fibril stretched to failure. We refer to the transition between the linear and non-linear behaviour as the fibril yield stretch/strain because there is evidence that stretching beyond this point leads to the accumulation of damage in the fibril \cite{shen2008stress}. The fibril yield strain can vary considerably between fibrils extracted from different sources. It has been reported to be approximately 6\% strain in groups of rabbit patellar tendon fibrils \cite{miyazaki1999tensile}, and 12\% in single rat patellar tendon fibrils \cite{liu2016tension}. In fibrils extracted from sea cucumber dermis, the yield strain shows a large amount of variation, falling anywhere between 6--55\% strain \cite{shen2010vitro}. After yielding and experiencing a decrease in modulus, the fibrils rupture at anywhere from 7\%  strain \cite{yamamoto2017relationships} to over 100\% strain \cite{shen2010vitro}. We can conclude that there is not a single value of fibril yield/rupture strain, but rather there is a distribution of these strains present in any given tendon. We also know that the structure of fibrils in tendons is not uniform. For example, the fibril diameter follows a trimodal distribution \cite{chang2020circadian}. It is possible that these mechanical and structural properties are related. 

Ideally, a microstructural model will only contain parameters that can be measured experimentally. This allows the model to predict how certain processes, such as ageing, disease, or injury, may affect macroscale tendon behaviour, provided we know how the microstructural components are affected. For example, in the tendons of patients with Ehlers-Danlos syndrome (EDS), the distribution of collagen fibril diameters is disrupted by a reduced quantity of collagen V \cite{sun2015targeted}, resulting in a diameter distribution with increased broadness and a larger mean diameter. A microstructural model that incorporates fibril diameter dependence could therefore be used to predict how the macroscale tendon properties would differ in comparison to a patient without EDS. This approach would be particularly useful in instances where the effects of a disease on the mechanical properties of the tendon are not clear, as it would allow some properties, such as fibril diameter distribution, to be held constant whilst others, such as fibril density, are varied.

Continuum damage models are frequently used to predict the post-yield behaviour of tendons as they are stretched to failure, although they often contain parameters that are not based on the microstructure. Natali et al. \cite{natali2003transversally,natali2005anisotropic} published a model where the strain energy function of the tendon is split into two parts associated with the matrix and fibrils, with each part being premultiplied by a damage function. The fibril damage function was derived by considering the number of fibrils which had yielded, assuming that the critical stretch required to remove crimp from the fibrils is normally distributed. The strain energy function used for the elastic regime, however,  was phenomenological and not related to the mechanical behaviour of individual collagen fibrils. Similar models were published by by Rodríguez et al. \cite{rodriguez2006stochastic} and Alastrué et al. \cite{alastrue2007structural}, where the behaviour of collagen fibrils is described by models based on polymer mechanics. Whilst these models are capable of predicting the general behaviour of tendons stretched to failure, they are all phenomenological to some degree and, consequently, they contain parameters that cannot be directly measured. Other models have used a microstructural approach, but were limited to modelling regions I and II of the stress-strain curve \cite{shearer2015a,shearer2015b,shearer2020recruitment}. 
 
Extending the widely used collagen recruitment model of Lanir \cite{lanir1983constitutive}, Hurschler et al. \cite{hurschler1997structurally}  developed an approach to model past the end of the linear region by including a yield criterion for the fibrils. By excluding both crimped and ruptured fibrils from stress calculations, Hurschler et al. \cite{hurschler1997structurally} were able to get reasonable fits to data by assuming the fibril critical stretch follows a Weibull distribution. Hamedzadeh et al. \cite{hamedzadeh2018constitutive} independently arrived at the same model, but extended its applicability by allowing the tissue to be compressible. They also showed how to model the effects of repeated overloading, demonstrating that it is possible to predict hysteresis whilst ignoring viscoelastic effects. The models of Hurschler et al. and Hamedzadeh et al., which we shall refer to after their authors as the HLV and HGF models, respectively, show that it is possible to model whole tendon behaviour as it is stretched to failure by only focussing on the failure behaviour of the fibrils. 

In this paper we use stress-strain data from Goh et al. \cite{goh2018age} to demonstrate the need for a new microstructural model of tendon failure. This stress-strain data was collected from failure tests carried out on mouse tail tendon fascicles, extracted from mice of different ages. The authors also provide structural data in the form of fibril diameter distributions, making it possible to explore the relationship between some of the structural and mechanical properties of tendons. In Section \ref{section:fitting existing models to data}, we attempt to fit a simplified version of the HLV and HGF models to this stress-strain data, showing that a significant proportion (47\%) of the data contains features that cannot be accounted for using these models.

In Section \ref{section: the elastic-plastic-distribution model}, we introduce a new model which is capable of capturing the range of stress-strain behaviour observed by Goh et al. \cite{goh2018age}. By using distributions to represent the fibril yield and rupture stretches, we demonstrate the range of stress-strain behaviour that can be generated by varying the shape of the distributions, and their position relative to one another. The resulting model includes only parameters that can, in principle, be measured directly, and can fit a wider range of stress-strain data than previous models once appropriate distributions have been selected. 
In section \ref{section: fitting the improved model to data}, we present some example fits to datasets with the features illustrated in Figure \ref{fig:introduction: plateau and steps}. We show that the parameter values found through fitting are consistent with those found experimentally.

\section{Fitting existing models to data}\label{section:fitting existing models to data}
\subsection{The elastic-rupture model}\label{subsection: the elastic-rupture model}
We first define the \textit{elastic-rupture} (ER) model, which is equivalent to the HLV model \cite{hurschler1997structurally}. We assume that the tendon is incompressible and composed of parallel, crimped fibres embedded within an isotropic matrix. We consider a simple uniaxial stretch $\lambda$ applied to the tendon, in the direction of the fibres, leading to a homogeneous stress field throughout the tissue. Each fibril has a critical stretch $\lambda_C$, which is the tendon stretch required to remove the crimp from the fibril. Once a fibril is taut, it exhibits a linear elastic response until it ruptures after being stretched by a factor of $\lambda_R$. By using a probability distribution $\Lambda_C(\lambda_C)$ to represent the variation of crimp found throughout the tissue, we can compute the stress in the tendon.

The shear modulus of the matrix has been estimated to be on the order of 1kPa \cite{shearer2017relative}, which is $\sim$1,000,000 times smaller than the fibril Young's modulus \cite{yamamoto2017relationships}. Assuming the matrix Young's modulus is of a similar magnitude, it is negligible compared to that of the fibrils, and we therefore choose to ignore any contributions to the stress from the matrix. We further assume that the deformation occurs at a strain rate that minimizes hysteresis, allowing us to ignore viscoelastic effects. Under these assumptions, the stress in the tendon is given by
	\begin{equation}\label{elastic-rupture tendon stress}
	\sigma_T^{ER}(\lambda) = \phi\int_{1}^{\lambda}\sigma_f^{ER}(\lambda,\lambda_C,\lambda_R)\Lambda_C(\lambda_C)\mathrm{d}\lambda_C.
	\end{equation}
where $\phi$ is the collagen volume fraction, and $\sigma_f^{ER}$ is the fibril stress. In the ER model, we define the fibril stress as
	\begin{equation}\label{elastic-rupture fibril stress}
	\sigma_f^{ER}(\lambda,\lambda_C,\lambda_R)=\left\{ \begin{aligned}
	&0, && \lambda < \lambda_C,\\ 
	&E\left(\frac{\lambda}{\lambda_C}-1\right), && \lambda_C \leq \lambda < \lambda_C\lambda_R, \\
	&0, && \lambda \geq \lambda_C\lambda_R,
	\end{aligned}
	\right.
	\end{equation}
where $E$ is the fibril Young's modulus. For the models used in this paper, we adopt a naming approach based on the physical behaviour of the fibrils. In the ER model, the fibrils are linearly elastic until they have been stretched by a factor of $\lambda_R$, after which they rupture. Throughout this paper, we will assume that the fibril critical stretch follows a triangular distribution, as in Hamedzadeh et al. \cite{hamedzadeh2018constitutive}, defined by
	\begin{equation}\label{triangular distribution definition}
	\Lambda_C(\lambda_C) = \left\{\begin{aligned}
	&0, &&\lambda_C< a,\\
	&\frac{2(\lambda_C-a)}{(b-a)(c-a)}, && a\leq\lambda_C<c,\\
	&\frac{2(b-\lambda_C)}{(b-a)(b-c)}, && c\leq\lambda_C<b,\\
	&0, && \lambda_C\geq b,
	\end{aligned}\right.
	\end{equation}
where $a$ is the lower limit, $b$ is the upper limit, and $c$ is the mode of the distribution. When using a triangular distribution for the fibril critical stretch, we can find an analytic expression for the tendon stress $\sigma_T^{ER}$. This expression can be found in \ref{appendix:ER analytic}.

\subsection{Fitting approaches}\label{subsection: fitting approaches}
We now fit the ER model to stress-strain data from Goh et al. \cite{goh2018age}. This data was gathered from mouse tail tendon fascicles, extracted from mice of different ages, which were stretched to failure. Along with stress-strain data, Goh et al. provide the tendon yield strain for each test specimen, and the mean collagen volume fraction for each age group. The authors defined the tendon yield strain to be the point with maximum gradient between the origin and the peak stress,	 after fitting a fifth order polynomial to the data. We explored two different fitting approaches, one which uses the yield stretch provided by Goh et al., and one which does not. In both cases we use the collagen volume fraction found by Goh et al. These two approaches are outlined below:
 
	\begin{enumerate}
	\item[] \textit{Generic fitting approach}: Fitting for all of the model parameters: $a$, $b$, $c$, $E$, and $\lambda_R$, using the whole range of stress-strain data (regions I--IV).
	\item[] \textit{Physically motivated fitting approach}: We assume $a=1$, meaning that fibrils immediately become taut upon stretching the tendon. We set the fibril rupture stretch $\lambda_R$ to be equal to the macroscopic yield point, provided by Goh et al. We then fit for the parameters $b$, $c$, and $E$ using the data in regions I and II. 
	\end{enumerate}
 
\begin{figure}[H]
	\centering
	\includegraphics[scale=1]{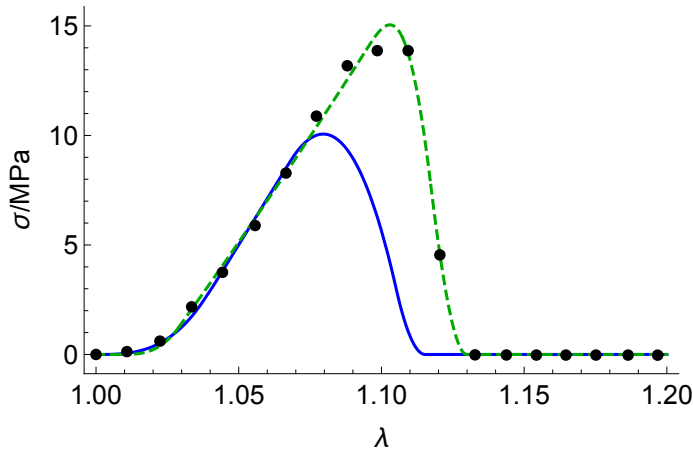}
	\caption{The ER model fitted to data from Goh et al. \cite{goh2018age}, using two different fitting approaches. The green dashed line was found using the generic fitting approach, whilst the solid blue line is the result of using the physically motivated fitting approach. The parameters from the generic fitting approach are: $E=6658$MPa, $a=1.008$, $b=1.107$, $c=1.095$, and $\lambda_R=1.020$. The parameters for the physically motivated fitting approach are: $E=386.7$MPa, $a=1.0$, $b=1.044$, $c=1.035$, and $\lambda_R=1.068$. The collagen volume fraction for both fittings is $\phi=0.56$, which was taken from Goh et al. \cite{goh2018age}. The root mean squared error for the generic approach is 0.264MPa, whilst for the physically motivated approach it is 3.43MPa. The generic fitting approach provides a superior fit, but at the cost of unphysical parameter values.}
	\label{fig:fitting existing models to data: elastic-rupture model}
	\end{figure}

In both fitting approaches, we use the analytic form of the tendon stress (see \ref{appendix:ER analytic}) along with a non-linear least squares method to find the fitting parameters. When using the generic fitting approach, it was often the case that the parameter values found were unphysical. For example, in Figure \ref{fig:fitting existing models to data: elastic-rupture model}, $b>\lambda_R$, suggesting that some fibrils begin to rupture before all of the fibrils have become taut. Whilst this seems reasonable if there is a large range of fibril critical stretch values, it contradicts evidence from Hijazi et al. \cite{hijazi2019ultrastructural}, that damage only occurs in the fibrils once the entire tendon has been stretched beyond the end of the macroscopic linear region (i.e. $\lambda_R\geq b$). Furthermore, the generic fitting approach often leads to unrealistically high values of collagen fibril Young's modulus $E$. In Figure \ref{fig:fitting existing models to data: elastic-rupture model} we have $E=6658$MPa, which is significantly larger than the highest value we could find in any paper where isolated collagen fibrils have been stretched to failure ($1900\pm500$MPa for bovine achilles tendon fibrils under ambient conditions \cite{grant2008effects}). 

Using the generic fitting approach requires the entire range of stress-strain data (regions I--IV), in order to determine the parameters $a$, $b$, $c$, $E$, and $\lambda_R$. If we were only interested in modelling the elastic tendon behaviour (regions I--II), the ER model could be modified by changing the fibril stress, defined in equation \eqref{elastic-rupture fibril stress}, so that the fibrils never rupture ($\lambda_R = \infty$). Since this elastic version of the model is defined by the same set of parameters, excluding $\lambda_R$, we should be able to determine the elastic parameters ($a$, $b$, $c$, and $E$) with the data from regions I and II alone. In other words, we should not need to stretch a tendon to failure in order to determine the parameters that define the pre-yield portion of the stress-strain curve. The physically motivated fitting approach ensures that $\lambda_R\geq b$, and that the elastic parameters are determined using the elastic part of the stress-strain data alone. Figure \ref{fig:fitting existing models to data: elastic-rupture model} shows an example of these two fitting approaches on the same set of data.

The restrictions placed on the model's parameters when using the physically motivated fitting approach mean that in the vast majority of cases, it is not possible to get a good fit to the data in regions III and IV. The magnitude of the post-yield stress is consistently underestimated, as can be seen in Figure \ref{fig:fitting existing models to data: elastic-rupture model}, suggesting that there is something missing from the model. Although the ER model can provide a reasonable fit in certain cases when the generic fitting approach is used, we instead choose to modify the model so that we can still use the physically motivated fitting approach. This ensures that all fitting parameters are realistic, and that the elastic parameters are consistent with the values we would find if the tendon was not stretched to failure.

\subsection{Features that cannot be accounted for}\label{subsection: features that cannot be accounted for}
The ER model is only capable of describing the mechanical response of tendons in the cases where the stress-strain curve looks similar to the idealised response illustrated in Figure \ref{fig:introduction:tendon stress-strain}. In the stress-strain data from Goh et al. \cite{goh2018age}, a significant proportion of the data contains features that cannot be captured using the ER model, even when the constitutive behaviour of the fibrils is adjusted to more closely resemble experimental data. In this section we describe these features, discuss their prevalence, and demonstrate why a new model is required to capture them. A summary of the information presented in this section, split according to the age of the mouse from which the fascicle was extracted, can be seen in Table \ref{table:proportion of tests with unusual features} of \ref{appendix:classifying stress-strain data}.
 
\subsubsection{Additional linear regions}\label{subsubsection: additional linear regions}
The idealised stress-strain curve in Figure \ref{fig:introduction:tendon stress-strain} contains one linear region, in the elastic part of the response. Some sets of stress-strain data from Goh et al. \cite{goh2018age} also contain a second linear region, present after the tendon has yielded. The gradient of this second linear section can vary, but is always less than the gradient of the first linear section. In some cases we see a small decrease in gradient as the tendon yields and enters a second linear region, followed by a well-defined peak in the stress. In other cases, the gradient of the second linear region is close to zero, and the stress reaches a plateau rather than a well-defined peak. 

To determine whether a stress-strain curve contains a second linear region, we first isolate the data points before the peak. We then interpolate the data using splines over 50 equally spaced points, and look for groups of at least 10 interpolated data points where the gradient does not vary by over 10\% of the maximum global gradient. Using this approach we find that 86 of 260 sets of data ($\sim$33\%) contain a second linear region. 

Our analysis showed that the peak can vary dramatically in broadness. Data with well-defined peaks can often be fitted using the ER model, but when the peak is wider, the ER model fails to capture the post-yield behaviour. This may be due to the fact that the ER model does not incorporate fibril plasticity, leading to an underestimation in the magnitude of the post-yield stresses at the tendon scale. The failings of the ER model are most apparent in the 33\% of cases where there is a clear second linear region. Without adding fibril plasticity to the ER model, we cannot possibly achieve a plateau in the macroscale stress-strain curve.

\subsubsection{Step-like failure behaviour}\label{subsubsection: step-like failure behaviour}
The second feature that cannot be accounted for using the ER model is step-like failure behaviour in the macroscopic failure region. In the idealised stress-strain curve presented in Figure \ref{fig:introduction:tendon stress-strain}, the gradient of the stress in region IV begins at zero, decreasing smoothly until it reaches a minimum value, before increasing back to zero. This behaviour can be described using the ER model. In the experimental data from Goh et al. \cite{goh2018age}, some of the stress-strain curves seem to show steps in this region, where the second derivative of the stress changes sign multiple times. 

We classify a set of data as exhibiting step-like failure behaviour if there is at least one data point in region IV with a larger gradient than both of its neighbouring points. Using this criterion, we find that 54 of 260 sets of data ($\sim$21\%) contain step-like failure behaviour. This is a significant proportion of the data, further supporting the need for an improved model.

\section{The elastic-plastic-distribution model}\label{section: the elastic-plastic-distribution model}
\subsection{General framework}\label{subsection: general framework}
By making biologically-motivated adjustments to the ER model, we can begin to account for the stress-strain features described in Section \ref{subsection: features that cannot be accounted for}. The first of these is to modify the constitutive behaviour of the fibrils so that once they have been stretched by a factor of $\lambda_Y$, they yield and begin to undergo plastic deformation. The second is to introduce distributions for the fibril yield stretch $\lambda_Y$, and rupture stretch $\lambda_R$. We call the resulting model the \textit{elastic-plastic-distribution} (EPD) model. In the EPD model, we define the fibril stress to be
	\begin{equation}\label{elastic-plastic fibril stress}
	\sigma_f^{EPD}(\lambda,\lambda_C,\lambda_Y,\lambda_R)=\left\{ \begin{aligned}
	&0, && \lambda < \lambda_C, \\ 
	&E\left(\frac{\lambda}{\lambda_C}-1\right), && \lambda_C \leq \lambda < \lambda_C\lambda_Y, \\
	&p(\lambda,\lambda_C,\lambda_Y,\lambda_R), && \lambda_C\lambda_Y \leq \lambda < \lambda_c\lambda_R, \\
	&0, && \lambda \geq \lambda_C\lambda_R,
	\end{aligned} 
	\right. 
	\end{equation}
where $p(\lambda,\lambda_C,\lambda_Y,\lambda_R)$ is the plastic stress in a yielded fibril. Experimental evidence suggests that, when selecting a functional form for $p$, we should choose a ``flat" function which has a lower gradient than the initial linear portion of the fibril stress. We believe that in instances where the macroscale tendon stress is displaying a broad/flattened peak, the majority of fibrils are also exhibiting flattened stress-strain behaviour. 

We assume that the fibril critical stretch $\lambda_C$, yield stretch $\lambda_Y$, and rupture stretch $\lambda_R$, follow a multivariate distribution $\Lambda(\lambda_C,\lambda_Y,\lambda_R)$. We then find the stress in the tendon by integrating the fibril stress over this distribution,
	\begin{equation}\label{elastic-plastic-distribution tendon stress}
	\sigma_T^{EPD}(\lambda) = \phi\int_{-\infty}^{\infty}\int_{-\infty}^{\infty}\int_{1}^{\lambda}\sigma_f^{EPD}(\lambda,\lambda_C,\lambda_Y,\lambda_R)\Lambda(\lambda_C,\lambda_Y,\lambda_R)\mathrm{d}\lambda_C\mathrm{d}\lambda_Y\mathrm{d}\lambda_R.
	\end{equation}
\subsection{Simplifying assumptions}\label{subsection: simplifying assumptions}
In order to demonstrate how the EPD model can be used to produce the range of macroscale stress-strain curves observed experimentally, we make two simplifying assumptions. Firstly, we assume that the fibrils are bilinear elasto-plastic, choosing the following form for $p$,
	\begin{equation}\label{bilinear elastoplastic functional form of p}
	p(\lambda,\lambda_C,\lambda_Y,\lambda_R) = Ek(\lambda_Y-1) + E(1-k)\left(\frac{\lambda}{\lambda_C}-1\right),
	\end{equation}
where $k\in[0,1]$ is a factor describing the decrease in gradient after the fibril has yielded. Secondly, we assume that the critical, yield, and rupture stretches are independent of each other, so that the distribution $\Lambda(\lambda_C,\lambda_Y,\lambda_R)$ can be written as
	\begin{equation}\label{independent stretch distributions}
	\Lambda(\lambda_C,\lambda_Y,\lambda_R) = \Lambda_C(\lambda_C)\Lambda_Y(\lambda_Y)\Lambda_R(\lambda_R).
	\end{equation}
We further assume that $\Lambda_C(\lambda_C)$, $\Lambda_Y(\lambda_Y)$, and $\Lambda_R(\lambda_R)$ are all triangular distributions. By changing the values of the parameters defining the yield and rupture stretch distributions, we can control the width of the macroscale stress-strain curve in regions III and IV. This amount of control is not possible using the ER model, and is required in order to fit the range of data observed experimentally.

\subsection{Marginal distributions} \label{subsection: marginal distributions}
In the HGF model \cite{hamedzadeh2018constitutive}, the authors define a damage distribution by stretching the critical stretch distribution by a factor of $\lambda_R$, their single value of rupture stretch. This damage distribution describes the proportion of fibrils in the tendon that have failed for a given value of tendon stretch. We can also compute equivalent distributions for the yield and rupture stretch when we have distributions, rather than single values. 

A fibril with critical stretch $\lambda_C$, and yield stretch $\lambda_Y$, will yield when the tendon stretch is equal to $\lambda=\lambda_C\lambda_Y$. We consider the joint distribution of critical stretch and yield stretch, and use this relation to define the following function
	\begin{equation}
		g(\lambda_C, \lambda) = \Lambda_C(\lambda_C)\Lambda_Y\left(\frac{\lambda}{\lambda_C}\right).
	\end{equation}
We then define $\overline{\Lambda_Y}(\lambda)$ as the marginal distribution found by integrating $g(\lambda_C,\lambda)$ with respect to $\lambda_C$,
	\begin{align}\label{marginal yield distribution}
	\overline{\Lambda_Y}(\lambda) = \int_{-\infty}^{\infty} g(\lambda_C,\lambda)\mathrm{d}\lambda_C = \int_{-\infty}^{\infty} \lambda_C(\lambda_C)\Lambda_Y\left(\frac{\lambda}{\lambda_C}\right)\mathrm{d}\lambda_C.
	\end{align}
We can follow a similar process in order to determine the marginal rupture distribution $\overline{\Lambda_R}(\lambda)$. The resulting distributions, once normalised, can be used to describe the proportion of fibrils that have yielded or ruptured for a given tendon stretch $\lambda$. Hamedzadeh et al. \cite{hamedzadeh2018constitutive} show the tendon stress-strain curve obtained using their model when the damage distribution (marginal rupture distribution in our model) and critical stretch distribution overlap. If these distributions overlap, then the first fibril may fail before the last fibril becomes taut, a scenario we argued against when justifying the physically motivated fitting approach. It is possible, however, that the marginal \textit{yield} distribution and the marginal rupture distribution can overlap in our model. As discussed in Section \ref{section:introduction}, there is a large range of yield and rupture strain values reported in the literature, even for fibrils extracted from the same source. It is therefore possible that as a tendon is stretched to failure, some fibrils rupture before others have yielded. 

\subsection{Varying the stretch distributions}	\label{subsection: varying the stretch distributions}
\begin{figure}[H]
	\centering
	\begin{subfigure}[H]{0.5\textwidth}
		\centering
		\includegraphics[width=\textwidth]{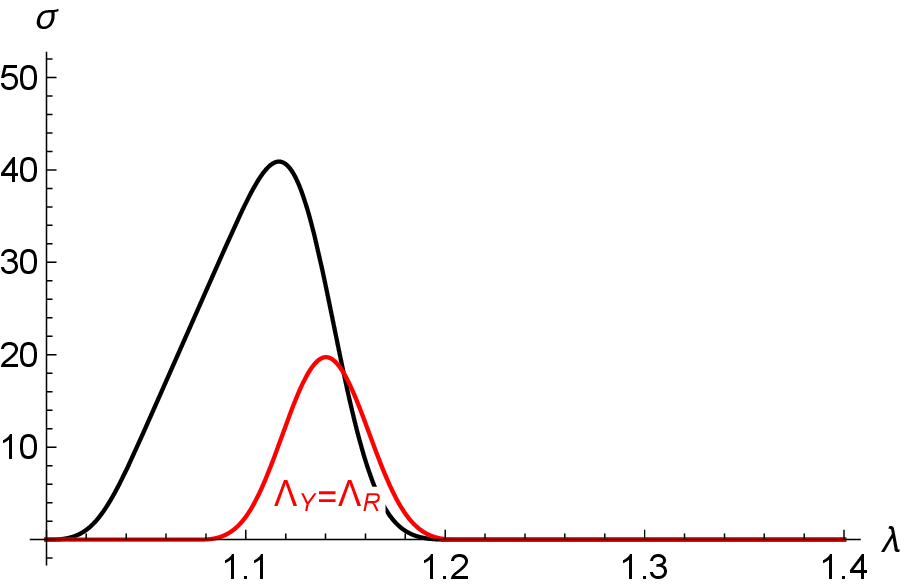}
		\caption{$\mu_R-\mu_Y = 0$}
		\label{sep=0}
	\end{subfigure}%
	~ 
	\vspace{1cm}
	\begin{subfigure}[H]{0.5\textwidth}
		\centering
		\includegraphics[width=\textwidth]{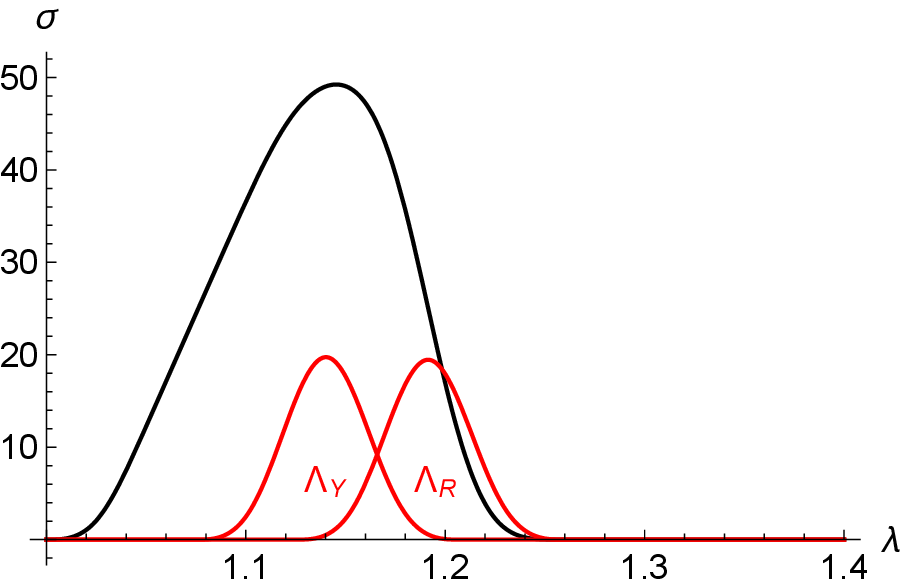}
		\caption{$\mu_R-\mu_Y = 0.05$}
		\label{sep=0.05}
	\end{subfigure}
	\begin{subfigure}[H]{0.5\textwidth}
		\centering
		\includegraphics[width=\textwidth]{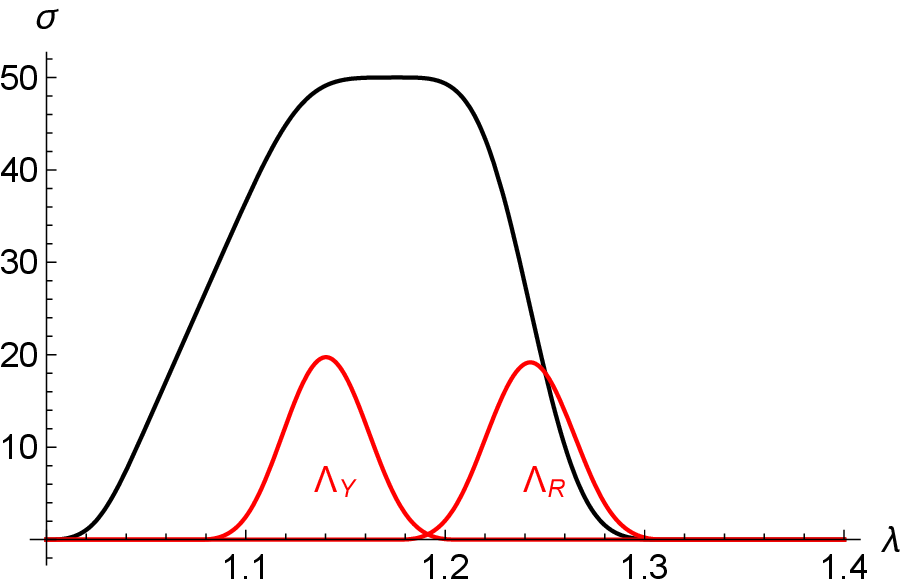}
		\caption{$\mu_R-\mu_Y = 0.1$}
		\label{sep=0.1}
	\end{subfigure}%
	~ 
	\begin{subfigure}[H]{0.5\textwidth}
		\centering
		\includegraphics[width=\textwidth]{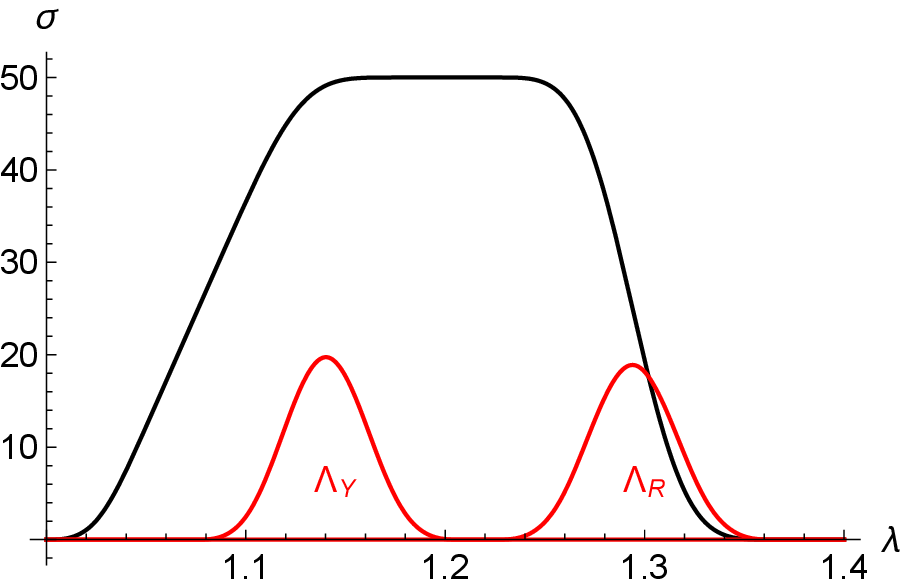}
		\caption{$\mu_R-\mu_Y = 0.15$}
		\label{sep=0.15}
	\end{subfigure}
	\caption{The effects of varying the separation between the yield stretch and rupture stretch distributions, on the macroscale stress-strain curve. The yield distribution was fixed and the position of the rupture distribution was varied. Symmetric triangular distributions were used for all of the stretch distributions. The black curves show the stress in the tendon, whilst the red curves show the marginal yield and rupture distributions used to generate them. The mean value of the (original, not marginal) yield and rupture distributions are $\mu_Y$ and $\mu_R$, respectively. $\mu_Y$ remains fixed at $\mu_Y = 1.1125$. We choose a value of $k=0$ to illustrate how varying the separation between the two distributions can lead to a plateau in the tendon stress.}
	\label{fig:general model: changing separation}
	\end{figure}

By varying the shape, position, and spread of the yield stretch and rupture stretch distributions, it is possible to produce macroscale stress-strain curves with the full range of features observed experimentally. Figure \ref{fig:general model: changing separation} shows the effects of varying the separation between the distributions, with all other parameters fixed.

\begin{figure}[H]
	\centering
	\begin{subfigure}[t]{0.5\textwidth}
		\centering
		\includegraphics[width=\textwidth]{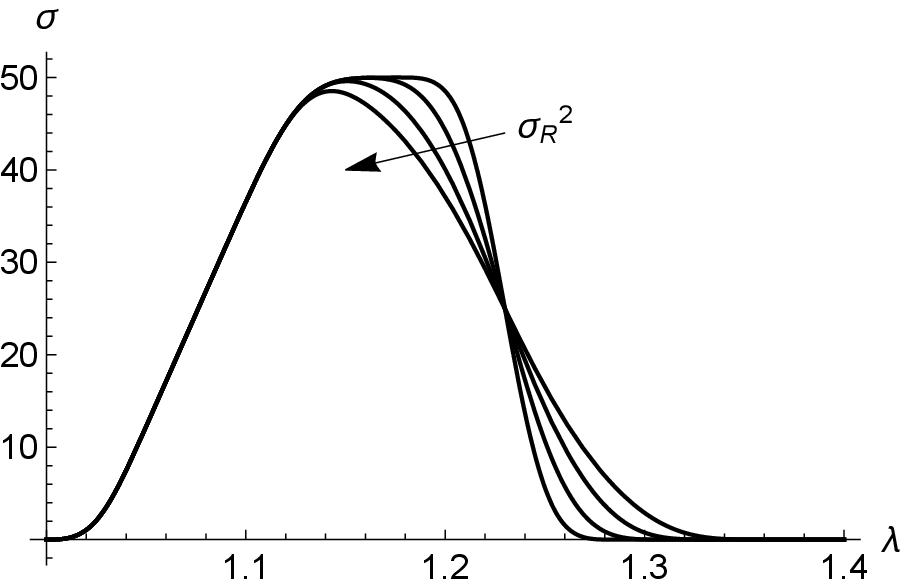}
		\caption{}
		\label{stress spreading rupture}
	\end{subfigure}%
	\begin{subfigure}[t]{0.5\textwidth}
		\centering
		\includegraphics[width=\textwidth]{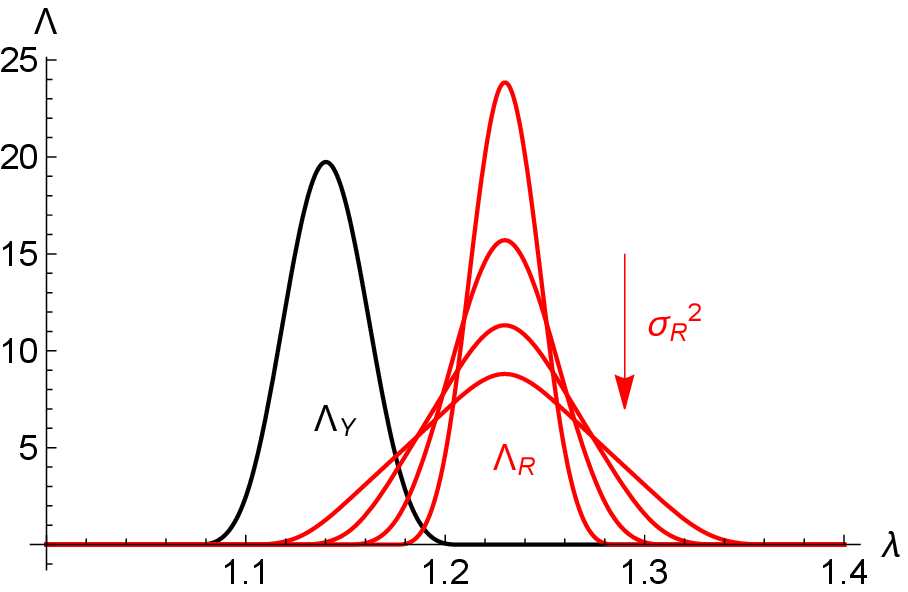}
		\caption{} 
		\label{distribution spreading rupture}
	\end{subfigure}
	\caption{When the mean values of the yield and rupture distributions are held constant and the variance of the rupture distribution is increased, the macroscale stress strain-curve changes as shown in (a). The corresponding marginal distributions are shown in (b). The arrows point in the direction of increasing variance, $\sigma_R^2$.}
	\label{fig:general model: rupture distribution spreading}
	\end{figure}

\begin{figure}[H]
	\centering
	\begin{subfigure}[t]{0.5\textwidth}
		\centering
		\includegraphics[width=\textwidth]{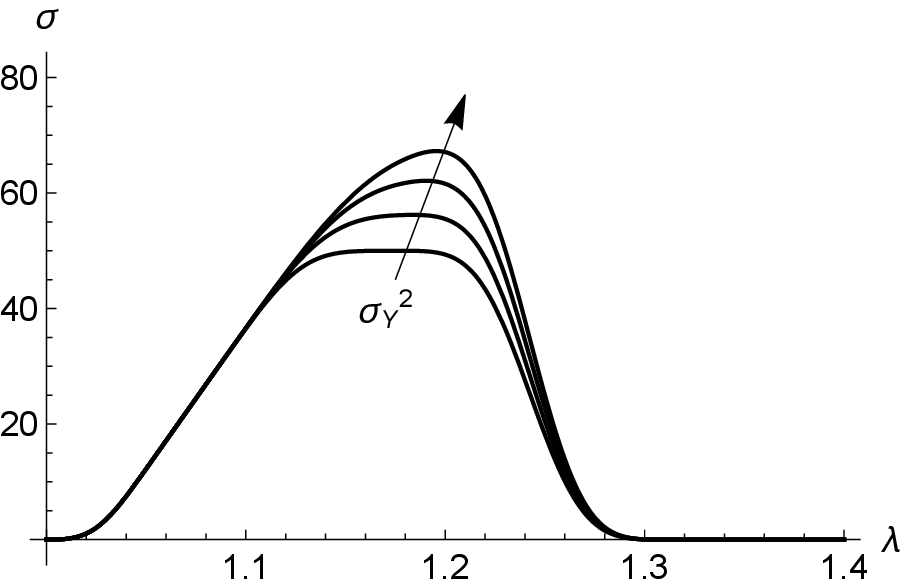}
		\caption{}
		\label{stress spreading yield}
	\end{subfigure}%
	\begin{subfigure}[t]{0.5\textwidth}
		\centering
		\includegraphics[width=\textwidth]{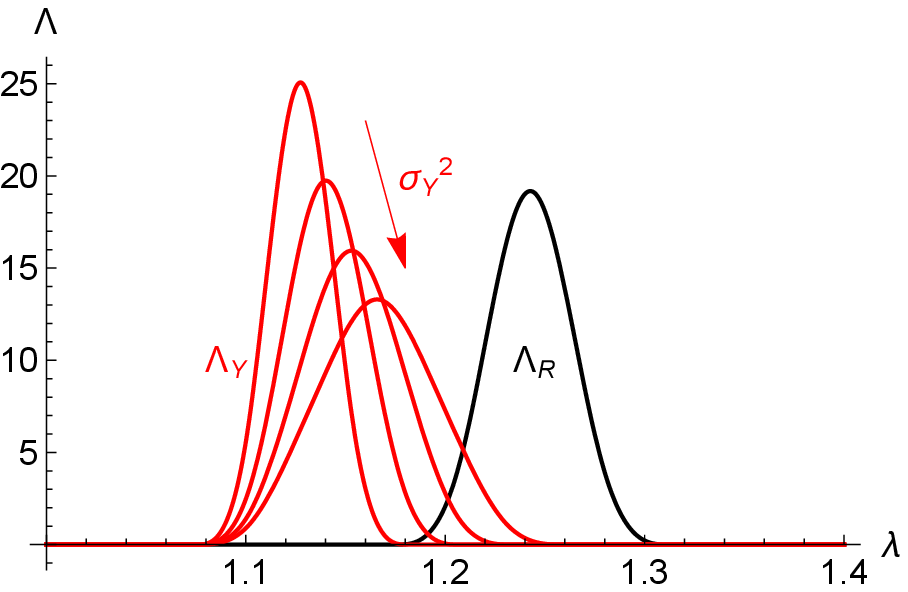}
		\caption{}
		\label{distribution spreading yield}
	\end{subfigure}
	\caption{Shown in (a) is the macroscale stress-strain curve obtained when the rupture stretch distribution is held constant and the variance of the yield stretch distribution is changed. The marginal distributions that generate these curves can be seen in (b). The variance is changed by increasing the upper limit of the distribution, whilst holding the lower limit constant. This is done so that the macroscale yield point remains constant in all of the stress-strain curves in (a). The arrows point in the direction of increasing variance, $\sigma_Y^2$. }
	\label{fig:general model: yield distribution spreading}
	\end{figure}

When the distributions overlap, there is a well-defined peak with no second linear region. Increasing the separation between the yield and rupture distributions causes a plateau to appear in the macroscale stress-strain curve, for values of tendon stretch between the marginal distributions. In this region, when there is no overlap, all of the fibrils are deforming plastically. By choosing the constitutive behaviour of the fibrils to be bilinear elasto-plastic, this leads to a second linear region. The spread of the yield and rupture distributions also affects the macroscale stress-strain curve, as can be seen in Figures \ref{fig:general model: rupture distribution spreading} and \ref{fig:general model: yield distribution spreading}.

In Figure \ref{fig:general model: rupture distribution spreading} we see that changing the spread of the rupture distribution, whilst holding the yield distribution constant, changes the shape of the peak and increases the width of region IV. As the spread increases, there is more overlap, and the peak becomes sharper. This is due to the fibrils rupturing sooner, and therefore no longer contributing to the macroscale stress. Figure \ref{fig:general model: changing separation} shows how translating the rupture stretch distribution to the right can result in a significant increase to the magnitude of the post-yield stress. Varying the spread of the rupture distribution, however, changes the shape of the peak without causing a significant increase in the magnitude of the stress in region III.

The results displayed in Figure \ref{fig:general model: yield distribution spreading} demonstrate how the macroscale stress-strain curve is affected when the rupture distribution is fixed and the spread of the yield distribution is varied. In doing this we choose to fix the lower bound of the yield distribution so that the macroscale yield point remains the same. By increasing the spread in the yield distribution, we are delaying the yielding of fibrils, causing an increase in the magnitude of the post-yield stress. We also increase the amount of overlap between the marginal distributions, leading to a sharper peak. The stress in the macroscopic failure region (region IV) is not affected by these changes.

\begin{figure}[H]
	\centering
	\begin{subfigure}[H]{0.5\textwidth}
		\centering
		\includegraphics[width=\textwidth]{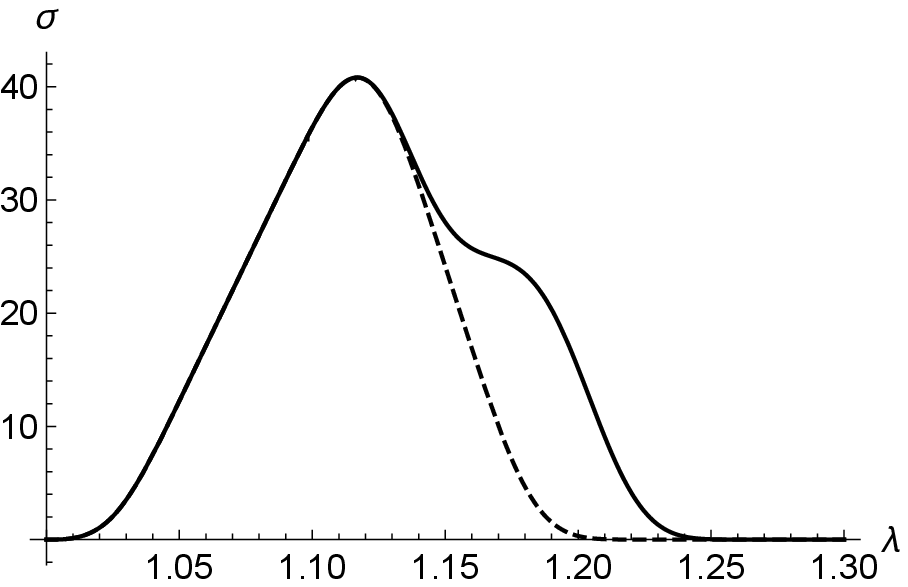}
		\caption{}
		\label{multi modal stress}
	\end{subfigure}%
	\begin{subfigure}[H]{0.5\textwidth}
		\centering
		\includegraphics[width=\textwidth]{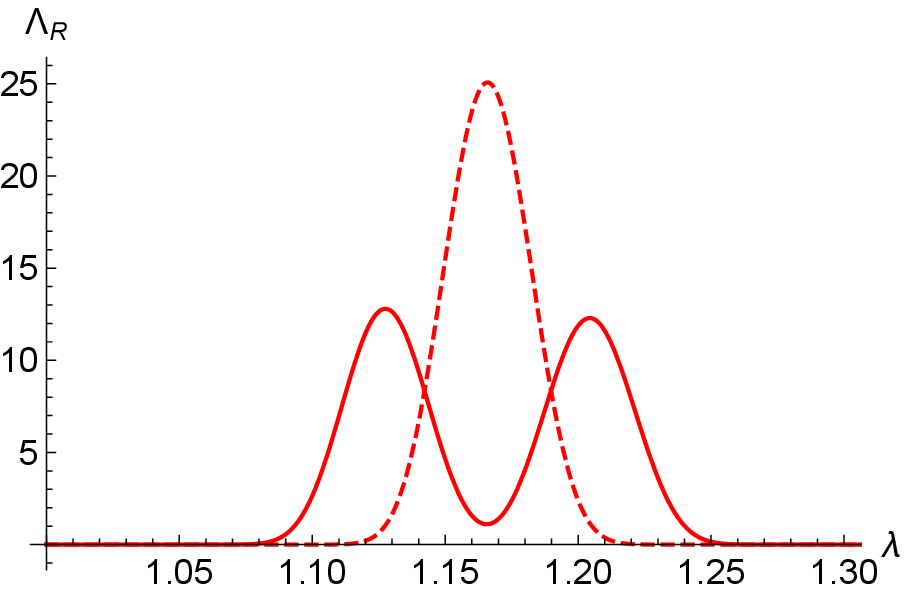}
		\caption{}
		\label{multi modal distributions}
	\end{subfigure}
	\caption{Changing the fibril rupture distribution from unimodal to bimodal causes the stress in the macroscopic failure region to exhibit step-like behaviour. Shown in (a) is the macroscopic stress-strain curve generated from the rupture distributions whose marginal counterparts are shown in (b). Dashed and solid lines correspond to the unimodal distribution and bimodal distribution, respectively. In the bimodal case, fewer fibrils are rupturing when the tissue stretch $\lambda$ falls between the modes, leading to a region in the macroscale stress-strain curve where the gradient becomes less steep.}
	\label{fig:general model: mutlimodal rupture distribution}
	\end{figure}

The distributions used to generate the stress-strain curves in Figures \ref{fig:general model: changing separation}, \ref{fig:general model: rupture distribution spreading}, and $\ref{fig:general model: yield distribution spreading}$ have all been unimodal triangular distributions. Allowing the rupture distribution to be multimodal causes the stress in the macroscopic failure region to exhibit step-like behaviour, as shown in Figure \ref{fig:general model: mutlimodal rupture distribution}.

When the rupture distribution used to generate the macroscale stress-strain curve is multimodal and has a sufficiently wide gap between the peaks, we see a corresponding region of the stress-strain curve where the gradient of the stress becomes less steep. As the tendon stretch approaches the next peak, the gradient of the stress becomes more negative again as fibrils continue to fail and no longer contribute to the total stress. This leads to step-like behaviour, as shown in Figure \ref{multi modal stress}.

\section{Fitting the improved model to data}   \label{section: fitting the improved model to data}

In this section we use the physically motivated fitting approach to show that the EPD model can be used to generate realistic stress-strain curves, with microstructural parameters that fall within the range of values observed experimentally. 

We continue to use the stress-strain data from Goh et al. \cite{goh2018age}, using the following process to find appropriate parameter values:

	\begin{enumerate}
	\item We assume the fibril critical stretch $\lambda_C$ follows a triangular distribution, defined by the parameters $a$, $b$, and $c$, as described in equation \eqref{triangular distribution definition}.
	\item We use the method outlined in Section \ref{section:fitting existing models to data} to determine the elastic parameters and use the macroscopic yield point provided by Goh et al \cite{goh2018age}. To reduce the number of fitting parameters we assume that the critical stretch distribution is symmetric so that $c=(a+b)/2$.
	\item We assume that both the fibril yield stretch $\lambda_Y$, and fibril rupture stretch $\lambda_R$, follow symmetric triangular distributions, given by $\Lambda_Y$ and $\Lambda_R$, respectively.
	\item We label the yield stretch distribution parameters as $a_Y$ and $b_Y$, and the rupture stretch distribution parameters as $a_R$ and $b_R$. 
	\item We fix $a_Y$ to be equal to the macroscopic yield stretch, so that the linear region ends when the first fibril yields.
	\item We then use a nonlinear least squares method to determine the remaining four parameters. We fit for $b_Y-a_Y$, $a_R$, $b_R-a_R$, and $k$ using \verb|scipy.optimize.curve_fit| in Python 3. Fitting for differences between parameters ensures that the distribution parameters remain ordered.
	\end{enumerate}

	\begin{figure}[H]
	\centering
	\includegraphics[scale=1]{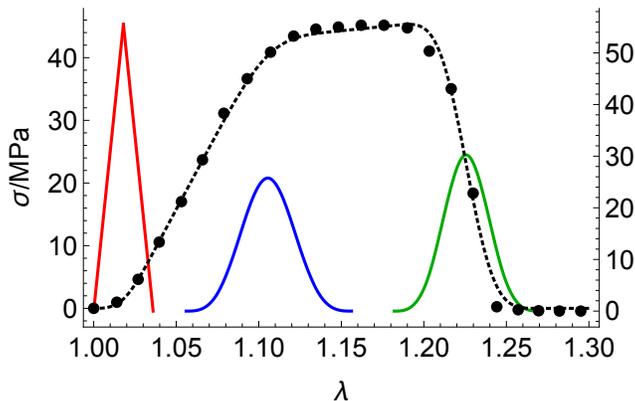}
	\caption{The black dashed line shows the EPD model, with bilinear elastoplastic fibrils, fitted to data from Goh et al \cite{goh2018age}. The critical stretch distribution, shown in red, was found by fitting the purely elastic model to the data in regions I and II. The blue and green curves show the marginal yield and rupture distributions, defined in equation \eqref{marginal yield distribution}. The right hand axis refers to these distributions. The critical, yield, and rupture stretch were assumed to follow symmetric triangular distributions, and the lower bound of the yield stretch distribution was fixed to correspond to the macroscopic yield point. The macroscale stress-strain curve contains a plateau region which could not be accounted for using previous models. The crimp distribution parameters are $a = 1.0$ and $b = 1.036$. The fibril Young's modulus is $E=895.7$MPa, and $k=0.053$. The yield stretch distribution parameters are $a_Y = 1.056$ and $b_Y = 1.116$. The rupture stretch distribution parameters are $a_R = 1.182$ and $b_R = 1.226$. The root mean squared error for the fit is 0.931MPa.}
	\label{fig:fitting EPD to data: plateau}
	\end{figure}

Figure \ref{fig:fitting EPD to data: plateau} shows an example of the EPD model fitted to data from Goh et al. \cite{goh2018age}, containing a plateau region rather than a well-defined peak. We can achieve a good fit to this data by assuming the yield stretch and rupture stretch follow symmetric triangular distributions. The root mean squared error for the fit is 0.931MPa, compared with 5.77MPa for the ER model using the generic fitting approach, and 25.0MPa for the ER model using the physically motivated fitting approach. The error is considerably larger for the ER model because it cannot generate a second linear region, and when using the physically motivated fitting approach, the stress goes to zero almost immediately after the tendon yields at $\lambda=1.056$. The initial non-linear toe region covers the support of the critical stretch distribution. The region between the supports of the critical stretch and marginal yield distributions corresponds to the macroscale linear region. Fibrils then begin to yield when the tendon stretch falls within the support of the marginal yield distribution, leading to a decrease in the gradient of the stress. There is then a region before fibrils start failing, where all fibrils are deforming plastically. The stress then decreases to zero as fibrils begin to rupture. 

	\begin{figure}[H]
	\centering
	\includegraphics[scale=1]{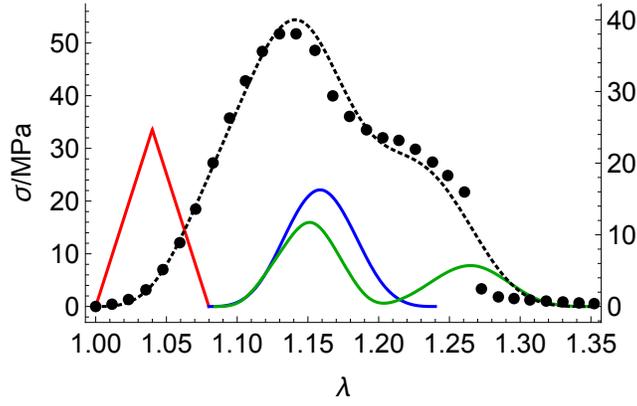}
	\caption{The EPD model fitted to data from Goh et al. \cite{goh2018age} containing a well-defined peak and step-like failure behaviour. The black dashed line shows the stress, whilst the red, blue, and green curves show the critical stretch, the marginal yield stretch, and the marginal rupture stretch distributions, respectively. The lower limit of the yield stretch distribution corresponds with the end of the macroscopic linear region, and due to the overlapping between the yield and rupture distributions, there is a well-defined peak in the stress strain curve. A bimodal rupture distribution is able to capture the step-like failure behaviour observed at the tendon level.  The crimp distribution parameters are $a=1.0$ and $b=1.081$. The fibril Young's modulus is $E=1160$MPa, and $k=0.064$. The yield stretch distribution parameters are $a_Y = 1.081$ and $b_Y=1.15$. The rupture stretch distribution parameters are $a_R^{(1)} = 1.085$, $b_R^{(1)} = 1.125$, $a_R^{(2)}= 1.180$, and $b_R^{(2)} = 1.253$. The second peak of the rupture stretch distribution has a weighting of $W=0.61$ relative to the first peak. The root mean squared error for the fit is 2.28MPa. }
	\label{fig:fitting EPD to data: steps}
	\end{figure}
Figure \ref{fig:fitting EPD to data: steps} shows a second set of data from Goh et al. \cite{goh2018age}, containing a well-defined peak along with step-like failure behaviour in region IV. The same fitting procedure was followed, but using a bimodal triangular distribution for the rupture stretch, in order to capture the step-like failure behaviour in region IV. We define a bimodal triangular distribution as
\begin{equation}
\Lambda_C^\text{bimodal}(\lambda_C) =\frac{1}{1+W}(\Lambda_C^{(1)}(\lambda_C) + W\Lambda_C^{(2)}(\lambda_C)),
\end{equation}
where $\Lambda_C^{(1)}$ and $\Lambda_C^{(2)}$ are unimodal triangular distributions, as defined in equation \eqref{triangular distribution definition}, with distribution parameters ($a^{(1)}$, $b^{(1)}$, $c^{(1)}$) and ($a^{(2)}$, $b^{(2)}$, $c^{(2)}$), respectively. The relative weighting between the two modes is given by $W$. 

The root mean squared error for the fit in Figure \ref{fig:fitting EPD to data: steps} is 2.28MPa, compared with 2.53MPa for the ER model using the generic fitting approach, and 18.3MPa for the ER model using the physically motivated fitting approach. The resulting yield and rupture stretch distributions overlap, leading to a well-defined peak in the stress. This also means that there is a range of tendon stretch values $\lambda$, where there simultaneously exists undamaged, yielded, and ruptured fibrils. The stress in the tendon begins to decrease as fibrils rupture, and once the tendon stretch passes into the region between the rupture distribution peaks, the gradient of the stress becomes less negative, because fewer fibrils are rupturing. As $\lambda$ passes into the second rupture distribution peak, the gradient becomes more negative, and the tendon eventually fails. 

\section{Discussion} \label{section: discussion}
We have shown that we can construct a mathematical model of tendon failure by splitting the fibril stress into elastic and plastic parts, and allowing the fibril yield stretch and rupture stretches to follow distributions, rather than being single-valued. When a single value of these parameters is used, as in the ER model, the plastic behaviour of the tendon is determined entirely by the critical stretch distribution. In at least 47\% of cases (see \ref{appendix:classifying stress-strain data}) it is not possible to get a good fit using the ER model because the stress-strain data contains a second linear region, step-like failure behaviour, or both. The EPD model provides a microstructural explanation for these features, and has the additional benefit of only including parameters that can be measured directly.

The introduction of a plastic stress function to the fibril constitutive behaviour is not new. Hamedzadeh et al. \cite{hamedzadeh2018constitutive} define their model in terms of a general fibril constitutive behaviour, thereby encompassing that part of the EPD model. The key difference in our approach is with the use of distributions to describe the fibril yield stretch and rupture stretch. Whilst modifying the fibril stress to become more flat can generate second linear regions, it is not enough to produce the range of behaviour observed by Goh et al \cite{goh2018age}. We must have a combination of both flattening fibril stress and a distribution of fibril yield stretch in order to capture all observed behaviour in region III. Without a distribution of yield stretch, the transition between a first and second linear region will be fixed by the fibril crimp distribution, meaning that sets of data with a narrow region I but a wide region III cannot be fitted using the model. By fixing the lower bound of the yield distribution to correspond with the macroscale tendon yield stretch, we can ensure that damage begins to occur in the fibrils when the macroscale linear region ends, as was documented by Hijazi et al. \cite{hijazi2019ultrastructural}. Without a distribution of fibril rupture stretch, we cannot produce step-like failure behaviour at the tendon scale without also incorporating it at the fibril scale. Failure tests carried out on individual collagen fibrils show that this would not be realistic. 

In our model, step-like failure occurs when one group of fibrils fails before another group. Yamamoto et al. \cite{yamamoto2017relationships} were able to determine a relationship between collagen fibril diameter and failure strain, for fibrils extracted from mouse tail tendons. Fibrils with a larger diameter seemed to fail at larger strains. It could be the case that in stress-strain data where we see step-like failure behaviour, there is a multimodal distribution of collagen fibril diameters, leading to a multimodal distribution of fibril failure strain. Although it is common to see multimodal fibril diameter distributions in tendons \cite{chang2020circadian}, the diameter distributions recorded by Goh et al. \cite{goh2018age} seem to be unimodal for mice in the age groups where step-like failure is most commonly observed (see Table \ref{table:proportion of tests with unusual features} in \ref{appendix:classifying stress-strain data}). Whilst this seems to contradict the theory that the step-like behaviour is due to groups of fibrils with different diameters failing in turn, it does not rule it out. The diameter distributions recorded by Goh et al. \cite{goh2018age} were found by taking the average across multiple fascicles, whilst the stress-strain data itself is from single fascicles. We cannot find the diameter distribution of a tendon and then stretch it to failure, as these are both destructive procedures. It remains plausible that the fascicles whose stress-strain curves contained step-like failure behaviour possess a multimodal diameter distribution, but that the average diameter distribution for that age group appears to be unimodal. Another possibility is that the relationship between collagen fibril diameter and rupture stretch is more complex than the linear relationship suggested by Yamamoto et al. \cite{yamamoto2017relationships}, somehow causing the unimodal diamater distribution to result in a multimodal rupture stretch distribution. It could also be the case that it is the distribution of rupture stretch that gives rise to the distribution of diameters as the tendon matures, i.e. the fibrils may grow differently depending on their initial mechanical properties.
We chose to present two examples of the EPD model fit to data, both of which contain features that cannot be accounted for with single values of fibril yield and rupture stretch. 
By assuming that the stretch distributions are symmetric, we are able to reduce the number of fitting parameters whilst still getting a good fit. In some cases it may be necessary to also fit for the mode $c$, or to use a different distribution entirely. We chose to use the macroscpic yield stretch found by Goh et al. \cite{goh2018age} as the lower bound of the fibril yield stretch distribution, meaning that any perceivable decrease in gradient corresponds to the accumulation of damage at the fibril scale. This leads to the upper bound of the crimp distribution in Figure \ref{fig:fitting EPD to data: steps} being equal to the macroscopic yield point, suggesting there is no purely linear region. When we model the fibrils as bilinear elasto-plastic, we only get a true linear region when the current tendon stretch $\lambda$ is not contained within the support of any of the marginal distributions. We can, however, get approximate linear regions over a larger range of tendon stretches if the value of the marginal distributions are small within that range. For example, in Figure \ref{fig:fitting EPD to data: plateau}, the flat region in the stress appears to be larger than the distance between the marginal yield and rupture distributions. 

It is difficult to judge whether the distributions used to generate the stress-strain curves in Figures \ref{fig:fitting EPD to data: plateau} and \ref{fig:fitting EPD to data: steps} are realistic because in all the references we could find, only a small number of collagen fibrils are stretched to failure. Liu et al. \cite{liu2016tension} state that for fibrils extracted from rat patellar tendons, the yield point falls between 10\% and 20\% strain. In Figure \ref{fig:fitting EPD to data: plateau}, the lower and upper bounds of the yield distribution are at 6\% strain and  12\% strain, respectively, whilst in Figure \ref{fig:fitting EPD to data: steps}, the yield distribution is bounded between 8\% and 14\% strain. Despite the data from Goh et al. \cite{goh2018age} and Liu et al. \cite{liu2016tension} coming from different sources, we believe that the similarities between the range of fibril yield strains observed experimentally, and the values we have found through our preliminary fitting, suggest that the EPD model is suitable for modelling the post-yield tendon behaviour (region III). 
 
Fibrils tested by Liu et al. \cite{liu2016tension} failed between 35\% and 107\% strain. Yamamoto et al. quote a failure strain of 34$\pm$11\% for fibrils extracted from mouse tail tendons \cite{yamamoto2017tensile}, and in a follow up paper, a range of 7--81\% \cite{yamamoto2017relationships}. All of these ranges exceed the upper bounds of the rupture distributions presented in Figures \ref{fig:fitting EPD to data: plateau} and \ref{fig:fitting EPD to data: steps}, which are 23\% and 25\%, respectively. Liu et al. \cite{liu2016tension} claim that because the tendon failure strain is typically much lower than the fibril rupture strain, there must be another component that is limiting the strength of the tendon, such as proteoglycans. We believe there are several reasons why the fibril rupture strain appears to be much larger than the tendon failure strain. Firstly, in tests carried out on individual collagen fibrils, the sections of fibril stretched to failure are much shorter than the entire length of the fibril. The fibrils tested by Yamamoto et al. \cite{yamamoto2017relationships} had a length of 19--64\textmu m, and whilst reliable data on the length of mouse tail tendon fibrils is not available, fibrils have been traced along the entire length (125\textmu m) of the mouse stapedius tendon \cite{svensson2017evidence}. If tail tendon fibrils are at least as long as those found in the stapedius tendon, then we can assume the sections tested by Yamamoto et al. were also significantly shorter than the total length of the fibril. Baldwin et al. \cite{baldwin2020new}  provide evidence that collagen fibrils, extracted from bovine tail, possess regions of mechanical susceptibility, due to a variation in structure along the length of the fibril. If fibrils are continuous throughout the length of the tendon, then the strength of the tendon will be limited by the strength of the weakest parts of its fibrils. By only testing small sections of fibrils to failure, rather than entire fibrils, these weakest sections are likely to be missed. This would lead to an overestimation of the fibril strength, potentially having a large effect on the perceived failure strain of collagen fibrils, depending on the frequency and strength of the regions of mechanical susceptibility. Secondly, it is possible that fibrils are subjected to inhomogeneous strains within the tendon. This could be due to the geometry of the tendon, or because the mechanical properties of the fibrils vary through the length of the tendon. An inhomogeneous strain applied to a fibril could cause it to rupture at a lower value of end-to-end strain than it would otherwise rupture at, outside of the tendon. Although some of the fibril failure strains reported in the literature \cite{shen2010vitro,yamamoto2017relationships} exceed the upper bounds of the rupture distributions used to generate the stress-strain curves in this paper, the quality of fit achieved demonstrates that the EPD model is still useful for modelling failure in tendons.
 
\section*{Acknowledgements}
Funding was provided by the UK Engineering and Physical Sciences Research Council (EPSRC) through the Doctoral Training Partnership (DTP).

\bibliography{tendon_failure}

\newpage
\appendix
\section{Analytic expression for the stress in the elastic-rupture model}\label{appendix:ER analytic}
When using the ER model, it is possible to determine an analytic expression for the stress in the tendon when the fibril critical stretch follows a triangular distribution. In this case, the tendon stress is given by

	\begin{equation}
	\sigma_{T}^{ER} (\lambda)= A^{ER} + B^{ER}\lambda + C^{ER}\lambda^2 + D^{ER}\lambda\log\lambda,
	\end{equation}

where $A^{ER}$, $B^{ER}$, $C^{ER}$, and $D^{ER}$ are piecewise constants defined by

	\begin{equation}\nonumber
	\mathclap{
	A^{ER}=\left\{ \begin{aligned}
	&0, &&\lambda<a &&&\text{and} &&&&\lambda/\lambda_R < a, \\
	&\frac{-a^2}{(a-b)(a-c)}, &&a\leq\lambda < c &&&\text{and} &&&&\lambda/\lambda_R <a, \\
	&\frac{-bc}{(a-b)(b-c)}-\frac{a}{a-b}, &&c\leq \lambda < b &&&\text{and} &&&&\lambda/\lambda_R <a, \\
	&-1,  &&\lambda \geq b &&&\text{and} &&&&\lambda/\lambda_R <a,\\
	&0, &&a\leq \lambda<c &&&\text{and} &&&&a\leq\lambda/\lambda_R<c,  \\
	&\frac{c^2}{(a-c)(c-b)}, &&c\leq \lambda<b &&&\text{and} &&&&a\leq\lambda/\lambda_R<c,  \\
	&\frac{b}{a-b}+\frac{ac}{(a-b)(a-c)}, &&\lambda\geq b &&&\text{and} &&&&a\leq\lambda/\lambda_R<c,  \\
	&0, &&c\leq \lambda < b &&&\text{and} &&&&c\leq\lambda/\lambda_R<b,  \\
	&\frac{b^2}{(a-b)(b-c)}, && \lambda \geq b &&&\text{and} &&&&c\leq\lambda/\lambda_R<b,  \\
	&0, &&\lambda \geq b &&&\text{and} &&&&\lambda/\lambda_R \geq b, 
	\end{aligned}\right. }
	\end{equation}

	\begin{equation}\nonumber
	\mathclap{
	B^{ER}=\left\{ \begin{aligned}
	&0, &&\lambda<a &&&\text{and} &&&&\lambda/\lambda_R < a, \\
	&\frac{2a\log a}{(a-b)(a-c)}, &&a\leq\lambda < c &&&\text{and} &&&&\lambda/\lambda_R <a, \\
	&\frac{2a\log a}{(a-b)(a-c)}+\frac{2c\log c}{(a-c)(b-c)}, &&c\leq \lambda < b &&&\text{and} &&&&\lambda/\lambda_R <a, \\
	&\frac{2a\log a}{(a-b)(a-c)} - \frac{2b\log b}{(a-b)(b-c)}+\frac{2c\log c}{(a-c)(b-c)},  &&\lambda \geq b &&&\text{and} &&&&\lambda/\lambda_R <a,\\
	&\frac{2a[\lambda_R(1-\log\lambda_R)-1]}{\lambda_R(a-b)(a-c)},  &&a\leq \lambda<c &&&\text{and} &&&&a\leq\lambda/\lambda_R<c,  \\
	&\frac{2c\log c}{(a-c)(b-c)} + \frac{2a[\lambda_R(1-\log\lambda_R)-1]}{\lambda_R(a-b)(a-c)}, &&c\leq \lambda<b &&&\text{and} &&&&a\leq\lambda/\lambda_R<c,  \\
	&\frac{-2b\log b}{(a-b)(b-c)}+\frac{2c\log c}{(a-c)(b-c)} + \frac{2a[\lambda_R(1-\log\lambda_R)-1]}{\lambda_R(a-b)(a-c)},  &&\lambda\geq b &&&\text{and} &&&&a\leq\lambda/\lambda_R<c,  \\
	&\frac{2b[\lambda_R(1-\log\lambda_R)-1]}{\lambda_R(a-b)(b-c)},  &&c\leq \lambda < b &&&\text{and} &&&&c\leq\lambda/\lambda_R<b,  \\
	&\frac{-2b\log b}{(a-b)(b-c)}+\frac{2b[\lambda_R(1-\log\lambda_R)-1]}{\lambda_R(a-b)(b-c)}, && \lambda \geq b &&&\text{and} &&&&c\leq\lambda/\lambda_R<b,  \\
	&0, &&\lambda \geq b &&&\text{and} &&&&\lambda/\lambda_R \geq b, 
	\end{aligned}\right. }
	\end{equation}
	
	\begin{equation}\nonumber
	\mathclap{
	C^{ER}=\left\{ \begin{aligned}
	&0, &&\lambda<a &&&\text{and} &&&&\lambda/\lambda_R < a, \\
	&\frac{1}{(a-b)(a-c)}, &&a\leq\lambda < c &&&\text{and} &&&&\lambda/\lambda_R <a, \\
	&\frac{1}{(a-b)(b-c)}, &&c\leq \lambda < b &&&\text{and} &&&&\lambda/\lambda_R <a, \\
	&0, &&\lambda \geq b &&&\text{and} &&&&\lambda/\lambda_R <a,\\
	&\frac{(\lambda_R-1)^2}{\lambda_R^2(a-b)(a-c)}, &&a\leq \lambda<c &&&\text{and} &&&&a\leq\lambda/\lambda_R<c,  \\
	&\frac{1}{\lambda_R^2(a-b)(a-c)}+\frac{1}{(a-b)(b-c)}-\frac{2}{\lambda_R(a-b)(a-c)}, &&c\leq \lambda<b &&&\text{and} &&&&a\leq\lambda/\lambda_R<c,  \\
	&\frac{(1-2\lambda_R)}{\lambda_R^2(a-b)(a-c)}, &&\lambda\geq b &&&\text{and} &&&&a\leq\lambda/\lambda_R<c,  \\
	&\frac{(\lambda_R-1)^2}{\lambda_R^2(a-b)(b-c)}, &&c\leq \lambda < b &&&\text{and} &&&&c\leq\lambda/\lambda_R<b,  \\
	&\frac{(1-2\lambda_R)}{\lambda_R^2(a-b)(b-c)},  && \lambda \geq b &&&\text{and} &&&&c\leq\lambda/\lambda_R<b,  \\
	&0, &&\lambda \geq b &&&\text{and} &&&&\lambda/\lambda_R \geq b, 
	\end{aligned}\right. }
	\end{equation}
	
	\begin{equation}\nonumber
	\mathclap{
	D^{ER}=\left\{ \begin{aligned}
	&0, &&\lambda<a &&&\text{and} &&&&\lambda/\lambda_R < a, \\
	&\frac{-2a}{(a-b)(a-c)}, &&a\leq\lambda < c &&&\text{and} &&&&\lambda/\lambda_R <a, \\
	&\frac{-2b}{(a-b)(b-c)}, &&c\leq \lambda < b &&&\text{and} &&&&\lambda/\lambda_R <a, \\
	&0, &&\lambda \geq b &&&\text{and} &&&&\lambda/\lambda_R <a,\\
	&0, &&a\leq \lambda<c &&&\text{and} &&&&a\leq\lambda/\lambda_R<c,  \\
	&\frac{-2b}{(a-b)(b-c)}+\frac{2a}{(a-b)(a-c)}, &&c\leq \lambda<b &&&\text{and} &&&&a\leq\lambda/\lambda_R<c,  \\
	&\frac{2a}{(a-b)(a-c)}, &&\lambda\geq b &&&\text{and} &&&&a\leq\lambda/\lambda_R<c,  \\
	&0, &&c\leq \lambda < b &&&\text{and} &&&&c\leq\lambda/\lambda_R<b,  \\
	&\frac{2b}{(a-b)(b-c)}, && \lambda \geq b &&&\text{and} &&&&c\leq\lambda/\lambda_R<b,  \\
	&0, &&\lambda \geq b &&&\text{and} &&&&\lambda/\lambda_R \geq b, 
	\end{aligned}\right. }
	\end{equation}

\section{Age breakdown of features present in stress-strain data}\label{appendix:classifying stress-strain data}
\begin{table}[H]
	\centering
	\begin{tabular}{@{}llllll@{}}
		\toprule
		Age group & $N_\text{total}$ & $N_\text{linear}$ & $N_\text{steps}$ & $N_\text{either}$ & $N_\text{both}$ \\ \midrule 
		1.6 month & 27 & 7 (26\%) & 16 (59\%) & 20 (74\%) & 3 (11\%) \\
		2.6 month & 25 & 13 (52\%) & 14 (56\%) & 19 (76\%) & 8 (32\%)  \\
		4.0 month  & 17 & 7 (41\%)  & 7 (41\%) & 11 (65\%)  & 3 (18\%) \\
		11.5 month & 34 & 16 (47\%) & 3 (9\%) & 18 (53\%) & 1 (3\%) \\
		23.0 month & 33 & 7 (21\%) & 3 (9\%) & 9 (27\%) & 1 (3\%) \\
		29.0 month & 43 & 11 (26\%) & 4 (9\%) & 14 (33\%) & 1 (2\%) \\
		31.5 month & 40 & 13 (33\%) & 3 (8\%) & 15 (38\%) & 1 (3\%) \\
		35.3 month & 41 & 12 (29\%) & 4 (10\%) & 15 (37\%) & 1 (2\%) \\
		\midrule 
		total:     & 260 & 86 (33\%) & 54 (21\%) & 121 (47\%) & 19 (7\%)\\ \bottomrule
	\end{tabular}
	\caption{The number of stress-strain curves from Goh et al. \cite{goh2018age} containing second linear regions $N_\text{linear}$, and step-like failure behaviour $N_\text{steps}$ for each age group of mice, as described in Section \ref{subsection: features that cannot be accounted for}. $N_\text{either}$ gives the number with either a second linear region, step-like failure behaviour, or both. The final column gives the number with both $N_\text{both}$.}
	\label{table:proportion of tests with unusual features}
\end{table}

\end{document}